\documentstyle[preprint,aps,eqsecnum]{revtex}
\draft
\newcommand{\ds}{\displaystyle}
\newcommand{\ul}{\underline}
\begin{document}
\title{Calculating excitation energies with the help of cumulants}
\author{Tom Schork and Peter Fulde}
\address{Max-Planck-Institut f\"ur Physik komplexer Systeme (Au\ss enstelle
Stuttgart), \\
Heisenbergstr.\ 1, 70569 Stuttgart, Germany.}
\date{\today}
\maketitle
\begin{abstract}
Recently, it has been shown that the ground-state energy of a quantum
many-body system can be written in terms of cumulants. In this paper we show
that the energies of excited states can be expressed similarly.  These
representations are suitable for various approximation schemes, e.g.,
projection techniques. The explicit use of cumulants ensures size consistency
in all approximations.

The theory is then applied to the computation of the energy bands of a
semiconductor with diamond structure. As we focus especially on the effects of
electron correlations, we consider a simplified model and re-derive results
obtained previously by the variational Local Ansatz method.
\end{abstract}

\section{Introduction}
Energy is a size-extensive quantity, i.e., the energy of two well-separated,
but otherwise identical systems equals twice the value of a single
system. Approximations which are applied when energies of a system are
calculated must preserve this property. In the standard diagrammatic approach
size consistency is ensured in any approximation by considering linked
diagrams only~\cite{NegeleOrland}.  However, the diagrammatic description
makes essential use of Wick's theorem, which is applicable only if the
dominant part of the Hamiltonian is a one-particle operator. Therefore this
approach is restricted to weakly correlated systems in which the
electron-electron interaction is weak and may be treated perturbatively.

It is known from classical statistical mechanics that size consistency is
attained by expressing size-extensive quantities in terms of
cumulants~\cite{UrsellMayer}. Kubo defined the cumulant of a quantum
mechanical expectation value~\cite{Kubo}. It is denoted by a superscript $^c$
and, with respect to the vector $\vert\phi\rangle$, it is given by
\begin{equation}
\langle\phi\vert A_1 \dots A_n \vert\phi\rangle^c =
\left. \frac{\partial}{\partial \lambda_1} \dots
\frac{\partial}{\partial \lambda_n} \log
\langle\phi\vert e^{\lambda_1 A_1} \dots e^{\lambda_n A_n}
\vert\phi\rangle \right|_{\lambda_1 = \dots = \lambda_n =0}
\label{cum}
\end{equation}
One immediately finds, e.g.,
\begin{equation}
\begin{array}{r c l}
\langle\phi\vert A_1 \vert\phi\rangle^c & = &
\langle\phi\vert A_1 \vert\phi\rangle
\\
\langle\phi\vert A_1 A_2 \vert\phi\rangle^c & = &
\langle\phi\vert A_1 A_2 \vert\phi\rangle -
\langle\phi\vert A_1 \vert\phi\rangle
\langle\phi\vert A_2 \vert\phi\rangle ~.
\end{array}
\end{equation}

Based on the Horn-Weinstein theorem~\cite{HornWeinstein}, an infinite series
termed ``connected moment expansion'' (CMX) has been derived for the
ground-state energy~\cite{CioslowskiI,Knowles,CioslowskiII}. Involving
cumulants only (rather than ordinary expectation values), any truncation of
the series leads to size-consistent estimates for the ground-state energy.

Here we would like to proceed as in Ref.~\cite{BeckerFulde}, where a different
representation of the ground-state energy in terms of cumulants has been
derived and which we refer to for details. The interacting many-body system is
described by a Hamiltonian $H$. We assume that the ground state
$\vert\Phi_0\rangle$ of a part $H_0$ of the Hamiltonian $H$ is known
\begin{equation}
H_0 \vert\Phi_0\rangle = \epsilon_0 \vert\Phi_0\rangle ~.
\end{equation}
Provided that $\vert\Phi_0\rangle$ has a finite overlap with the ground state
of $H$, the ground-state energy $E_0$ of $H$ is given by
\begin{equation}
E_0 = \left( H\vert \Omega \right)~.
\label{gsenergy}
\end{equation}
Here we introduced the abbreviation
\begin{equation}
(A \vert B) = \langle \Phi_0 \vert A^\dagger B \vert \Phi_0 \rangle^c
{}~.
\label{product}
\end{equation}
It serves as a form on the vector space (the Liouville space) spanned by the
operators which are defined on the usual Hilbert space. $\Omega$ is a
superoperator which acts on the Liouville space. It is defined by
\begin{equation}
\Omega = 1 + \lim_{z\to 0} \frac{1}{z-(L_0+H_1)} H_1 ~.
\label{omega}
\end{equation}
$L_0$ is the Liouville operator with respect to $H_0$. It acts on any operator
$A$ according to $L_0 A = [H_0, A]_-$. $\Omega$ characterizes the
transformation of the unperturbed ground state $\vert\Phi_0\rangle$ into the
ground state of $H$, $\vert\Psi_0\rangle$. Thus it bears similarities to the
wave operator of quantum mechanics and we will refer to it as ``wave
operator'' in the following. Note, however, that $\Omega$ appears only in an
expectation value, Eq.~(\ref{product}), and does not yield the wave function
$\vert\Psi_0\rangle$.

The explicit use of cumulants in Eq.~(\ref{gsenergy}) ensures size consistency
in any subsequent approximation and can be considered as replacing the
afore-mentioned ``linked diagrams''. However, it is important to notice that
the cumulant representation of the ground-state energy, Eq.~(\ref{gsenergy}),
does not have the limitations which the diagrammatic approaches often have
since it holds for an arbitrary decomposition of $H$ into $H_0$ and $H_1$. In
particular, $H_0$ need not be an one-particle Hamiltonian (e.g., Hartree-Fock)
as in the standard diagrammatic approach. Hence, the present description is
applicable to strongly correlated systems as well, e.g., to spin systems
described by the Heisenberg Hamiltonian.

Another advantage of using Eq.~(\ref{gsenergy}) is that quite different types
of approximations can be applied than in diagrammatic approaches. One example
is a partitioning of the Liouville space, which was pioneered by L\"owdin and
by the Uppsala and Gainsville Quantum Chemistry
Groups~\cite{Loewdin,Goscinski,BartlettBraendas}. Consider a set of operators
$A_\nu$, which span a subspace of the Liouville space. Taking into account
this ``relevant'' subspace only and, thereby, neglecting the remaining part
the following ansatz for the wave operator $\Omega$~(\ref{omega}) is made
\begin{equation}
\ds \Omega = 1+\sum_\nu \eta_\nu A_\nu ~.
\label{linansatz}
\end{equation}
The coefficients $\eta_\nu$ are determined by
\begin{equation}
(A_\nu\vert H\Omega) =0 ~.
\label{eqset}
\end{equation}
This equation holds generally for all operators $A$ and is easily proven by
inserting the definition~(\ref{omega}) of $\Omega$~\cite{SchorkFulde}. Making
the linear ansatz~(\ref{linansatz}) for $\Omega$ is equivalent to applying
projection technique to Eq.~(\ref{gsenergy})~\cite{BeckerFulde}. In
particular, if we choose the powers of the Hamiltonian as relevant operator
set, i.e., $A_\nu = H^\nu$ ($\nu=1\dots N$) we recover the improved CMX as in
Refs.~\cite{Knowles,CioslowskiII}. For a more general ansatz than
Eq.~(\ref{linansatz}) and other approximation schemes, see
Ref.~\cite{SchorkFulde}.

For these reasons it would be desirable to express not only the ground-state
energy, but also the excitation energies in terms of cumulants. The way this
has to be done can be conjectured by considering the Rayleigh-Schr\"odinger
perturbation theory. By expanding $\Omega$ in terms of $H_1$ and taking the
limit $z\to0$ one arrives at the perturbation expansion of the ground-state
energy:
\begin{equation}
E_0 = \epsilon_0 + \sum_{n=0}^{\infty} \left( H_1 \left\vert \left(
-\frac{1}{L_0} H_1 \right)^n \right.\right) ~.
\end{equation}
The generalization to excited states, which suggests itself, is to take the
expectation values with respect to an excited state $\vert\Phi_i\rangle$ of
$H_0$ instead of $\vert\Phi_0\rangle$
\begin{equation}
H_0 \vert\Phi_i\rangle = \epsilon_i \vert\Phi_i\rangle
\end{equation}
so that the expansion of the energy of an excited state of $H$ becomes
\begin{equation}
E_i = \epsilon_i + \sum_{n=0}^{\infty} \left( H_1 \left\vert \left(
-\frac{1}{L_0} H_1 \right)^n \right. \right)_i ~.
\label{rsi}
\end{equation}
The subscript $i$ indicates that the expectation values are now taken with
respect to the excited state $\vert\Phi_i\rangle$
\begin{equation}
(A \vert B)_i = \langle \Phi_i \vert A^\dagger B \vert \Phi_i
\rangle^c
\label{producti}
\end{equation}
[cf.\ Eq.~(\ref{product})]. Summing the geometric series in Eq.~(\ref{rsi}),
we arrive at the conjecture
\begin{equation}
E_i = \left( H \left| 1 + \lim_{z\to 0} \frac{1}{z-(L_0+H_1)} H_1
\right.\right)_i = \left( H\vert \Omega \right)_i
\label{conjecture}
\end{equation}
where $\Omega$ is defined as in Eq.~(\ref{omega}). Note the interchange of the
limits $z\to 0$ and $n\to\infty$.

In the next section, we show that the conjecture~(\ref{conjecture}) is indeed
true under certain circumstances. In Sec.~\ref{sec:application} we present a
specific application of the formalism by calculating the quasiparticle energy
bands of a semiconductor in the Bond-Orbital Approximation (BOA). In
Sec.~\ref{sec:discussion} we compare the results with previous work. Finally,
we give a brief summary and conclude in Sec.~\ref{sec:conclusion}.

\section{Expressing excitation energies in terms of cumulants}
\label{sec:theory}
In this section we prove that Eq.~(\ref{conjecture}) holds indeed provided
$\vert\langle\Phi_i\vert\Psi_i\rangle\vert^2 > 1/2$. Here,
$\vert\Psi_i\rangle$ denotes an eigenstate of $H$ with eigenvalue
$E_i$. Further, let
$\beta^{(i)}_\mu=\vert\langle\Phi_i\vert\Psi_\mu\rangle\vert^2$. We consider
the following function
\begin{equation}
f_i(t) =
\log \langle\Phi_i\vert e^{-it H} \vert\Phi_i\rangle
= \log \sum_\mu \beta^{(i)}_\mu e^{-it E_\mu}
\label{deffi}
\end{equation}
for $t\geq 0$. Note that the derivative of $f_i(-i\lambda)$ with respect to
$\lambda$ is the function introduced by Horn and Weinstein in
Ref.~\cite{HornWeinstein}, which tends to the ground-state energy at the limit
$\lambda\to\infty$ (provided that $\beta^{(i)}_0 \neq 0$).  However, the
function defined here need not tend to the ground-state energy as $t\to\infty$
since the arguments of the exponentials are imaginary. This is shown next.

The argument of the logarithm is a complex number and, therefore, we have to
specify the branch of the logarithm. This is done by requiring that $f_i(t)$
is analytic in a neighborhood of the positive real $t$-axis. As we assume that
$\beta^{(i)}_i > 1/2$, it is
\hbox{$\langle\Phi_i\vert\exp(-iHt)\vert\Phi_i\rangle \neq 0$} for all
$t\geq 0$ and this requirement can be fulfilled. We continue straightforwardly
\begin{equation}
\begin{array}{r c l}
f_i(t) & = & \ds \log \sum_\mu \beta^{(i)}_\mu e^{-itE_\mu} \\
& = & \ds\log \left[ \beta^{(i)}_i e^{-itE_i}
\left( 1+ \sum_{\mu\neq i} \frac{\beta^{(i)}_\mu}{\beta^{(i)}_i}
e^{-it(E_\mu-E_i)} \right) \right] \\
& = & \ds -it E_i + \log \beta^{(i)}_i + \log \left( 1+ \sum_{\mu\neq i}
\frac{\beta^{(i)}_\mu}{\beta^{(i)}_i} e^{-it(E_\mu-E_i)} \right)
+ 2\pi i n(t) ~.
\end{array}
\label{fi1}
\end{equation}
The integer function $n(t)$ guarantees that we remain on the correct branch of
the logarithm. $n(t)$ is fixed by the previous requirement, namely that
$f_i(t)$ is analytic for $t\geq 0$.  The first two terms of the last line in
Eq.~(\ref{fi1}) are analytic. The third term is analytic, too, because
\begin{equation}
\left|\sum_{\mu\neq i} \frac{\beta^{(i)}_\mu}{\beta^{(i)}_i}
e^{-it(E_\mu-E_i)}\right|
 \leq \sum_{\mu\neq i} \frac{\beta^{(i)}_\mu}{\beta^{(i)}_i} =
\frac{1}{\beta^{(i)}_i} - 1 < 1
\end{equation}
(remember that $\beta^{(i)}_i > 1/2$). Therefore $n(t)$ is a constant, which
vanishes because $n(0)=0$. As the argument of the logarithm in Eq.~(\ref{fi1})
is bounded the asymptotic behavior of $f_i(t)$ is
\begin{equation}
\frac{f_i(t)}{t} \to -iE_i \qquad (t\to\infty)~,
\label{asymptotic}
\end{equation}
i.e., the energy of the excited state is given by the behavior of $f_i(t)$ for
large $t$.

In order to calculate $E_i$ with the help of Eq.~(\ref{asymptotic}), we
consider the half-sided Fourier transform of $f_i$
\begin{equation}
\hat f_i(z)
= -i \int_0^\infty dt~ e^{itz} f_i(t) ~.
\label{fourier}
\end{equation}
Since $f_i(t)/t$ approaches a constant for $t\to\infty$, the integral is
well-defined for ${\rm Im}~z > 0$. From the asymptotic
behavior~(\ref{asymptotic}) we readily obtain the energy of the excited state
in terms of $\hat f_i(z)$:
\begin{equation}
E_i = \lim_{z\to i0^+} z^2 \hat f_i(z) ~.
\end{equation}

We finally introduce cumulants by making use of Eq.~(\ref{cum})
\begin{equation}
f_i(t) = \log \langle\Phi_i\vert e^{-iHt} \vert\Phi_i\rangle
= \langle\Phi_i\vert e^{-iHt}-1 \vert\Phi_i\rangle^c ~.
\end{equation}
This enables us to take the Fourier transform (\ref{fourier}) explicitly and
after some algebra (cf.\ Ref.~\cite{BeckerFulde}) we obtain
\begin{equation}
\hat f_i(z) = \frac{\epsilon_i}{z^2} + \frac{1}{z^2}
\left\langle\Phi_i\left\vert
H_1 \left(1+ \frac{1}{z-(L_0+H_1)}\right)
\right\vert\Phi_i\right\rangle^c ~.
\end{equation}
Hence the energy of the exited state is given by
\begin{equation}
E_i = \lim_{z\to i0^+} z^2 \hat f_i(z) = (H\vert \Omega)_i ~,
\label{theorem}
\end{equation}
where we again used the definition~(\ref{omega}) of $\Omega$ and the
abbreviation~(\ref{producti}).

\section{Application to semiconductors}
\label{sec:application}
The formalism presented in the previous section will now be used to
calculate the electronic excitation energies of a semiconductor. These
quasiparticle (quasihole) energies are defined as the energy
difference between an excited state with momentum $\ul{k}$ of the
$N+1$- ($N-1$-) particle system and the ground state of the
$N$-particle system, where $N$ denotes the number of electrons in the
charge neutral system.

To be specific, we treat a semiconductor of diamond-type structure.
It is described by orthogonalized atomic-like $sp^3$ hybrids and the
resulting functions are denoted by $g_i(\ul{r})$. The corresponding
electron creation (annihilation) operators $a_{i\sigma}^\dagger$
($a^{\phantom{\dagger}}_{i\sigma}$) fulfill the usual anticommutation
relations.

In this basis the Hamiltonian reads as follows:
\begin{equation}
\begin{array}{r c l}
H & = & H_0 + H_{\rm int} \\ H_0 & = & \ds\sum_{i,j,\sigma}
t^{\phantom{\dagger}}_{ij}
a_{i\sigma}^\dagger a^{\phantom{\dagger}}_{j\sigma}
\\
H_{\rm int} & = &
 \frac{1}{2} \ds \sum_{i,j,k,l,\sigma,\sigma'}
V^{\phantom{\dagger}}_{ijkl} a_{i\sigma}^\dagger a_{k\sigma'}^\dagger
a^{\phantom{\dagger}}_{l\sigma'} a^{\phantom{\dagger}}_{j\sigma} ~.
\end{array}
\end{equation}
Here
\begin{equation}
t_{ij} = \int d^3r ~ g_i^* (\ul{r})
\left( -\frac{1}{2m} \Delta + V_{\rm ext}(\ul{r}) \right)
g_j(\ul{r})
\end{equation}
is the bare hopping matrix element between the hybrid orbitals $i$ and
$j$, and $V_{\rm ext}(\ul{r})$ is the electrostatic potential set up
by the nuclei and the core electrons.
\begin{equation}
V_{ijkl} =
\int d^3 r d^3 r' ~ g_i^* (\ul{r}) g_j(\ul{r})
\frac{e^2}{|\ul{r}-\ul{r}'|} g_k^*(\ul{r}') g_l(\ul{r}')
\end{equation}
are the interaction matrix elements.

In the Hartree-Fock (HF) approximation, the effective one-particle
Hamiltonian is given by
\begin{equation}
H_{\rm HF} = \sum_{i,j,\sigma} f^{\phantom{\dagger}}_{ij}
a_{i\sigma}^\dagger a^{\phantom{\dagger}}_{j\sigma}
\label{HFHamil}
\end{equation}
where $f_{ij}$ denotes the Fock matrix:
\begin{equation}
f_{ij} = t_{ij} + \sum_{k,l,\sigma'}
\left( V_{ijkl} - \frac{1}{2} V_{ilkj} \right)
\langle\Phi^{\rm HF}\vert a^\dagger_{k\sigma'}
a^{\phantom{\dagger}}_{l\sigma'} \vert\Phi^{\rm HF}\rangle ~.
\label{Fockmatrix}
\end{equation}
$\vert\Phi^{\rm HF}\rangle$ is the HF ground state which has to be
determined self-consistently from Eqs.~(\ref{HFHamil}) and
(\ref{Fockmatrix}). The HF ground-state energy is given by
\begin{equation}
E^{\rm HF}_0 = \langle\Phi^{\rm HF}\vert H_0 + H_{\rm int}
\vert\Phi^{\rm HF}\rangle + E_{\rm NN}
\end{equation}
where $E_{\rm NN}$ denotes the interaction energy of the nuclei and
the core electrons.

To proceed further, we introduce bonding and antibonding functions for
each bond $I$ with corresponding creation operators
\begin{equation}
\begin{array}{r c l}
B^\dagger_{I\sigma} & = & \ds
\frac{1}{\sqrt 2} ( a^\dagger_{I1\sigma} + a^\dagger_{I2\sigma}) \\
A^\dagger_{I\sigma} & = & \ds
\frac{1}{\sqrt 2} ( a^\dagger_{I1\sigma} - a^\dagger_{I2\sigma}) ~.
\end{array}
\end{equation}
The operators $a^\dagger_{I1\sigma}$ and $a^\dagger_{I2\sigma}$ refer
to the $sp^3$ hybrids $I1$ and $I2$ which form bond $I$. In the
bond-orbital approximation (BOA)~\cite{Harrison} the $N$-particle HF
ground state is approximated by occupying all bonding states
\begin{equation}
\vert\Phi^{\rm HF}\rangle =
\prod_{I,\sigma} B^\dagger_{I\sigma} \vert 0 \rangle ~.
\end{equation}
In a more refined calculation the bonding orbitals are constructed
from the HF orbitals by applying the localization method of
Foster and Boys \cite{FosterBoys} or with the help of a localization
potential, which acts within the space of the occupied HF orbitals
only.

When an electron with momentum $\ul{k}$ is added to the system the
$N+1$-particle HF wave function is given within the BOA by
\begin{equation}
\vert \Phi^{\rm HF}_{\ul{k}m\sigma} \rangle
= A_{\ul{k}m\sigma}^\dagger \vert \Phi^{\rm HF} \rangle
\label{HFfunctionk}
\end{equation}
where $A_{\ul{k}m\sigma}^\dagger$ creates an electron with momentum
$\ul{k}$ in the conduction band $m$. The operator
$A_{\ul{k}m\sigma}^\dagger$ is decomposed into
\begin{equation}
A^\dagger_{\ul{k}m\sigma} = \sum_{I}
\alpha^{\phantom{\dagger}}_{\ul{k}m\sigma}(I) A^\dagger_{I\sigma}
\end{equation}
and the $\alpha_{\ul{k}m\sigma}(I)$ are found from the solution of the
HF eigenvalue problem.

The wave function~(\ref{HFfunctionk}) yields an energy eigenvalue
$E^{\rm HF}_{\ul{k}m\sigma}$. From this we find the energy of a
quasiparticle in the conduction band
\begin{equation}
\epsilon^{\rm HF}_{\ul{k}m\sigma} =
E^{\rm HF}_{\ul{k}m\sigma} - E^{\rm HF}_0 ~,
\end{equation}
which, according to Koopmans' theorem~\cite{Koopmans}, coincides with
an eigenvalue of the one-electron Hamiltonian $H_{\rm HF}$. Similarly,
the valence bands are calculated.

We now want to determine the correlation contributions to these
quasiparticle energies. They reflect the influence of the residual
Hamiltonian
\begin{equation}
\begin{array}{r c l}
H_{\rm res} & = & H-H_{\rm HF} \\
& = & H_{\rm int} -
 \frac{1}{2} \ds\sum_{i,j,k,l,\sigma,\sigma'}
(V_{ijkl}-\delta_{\sigma,\sigma'} V_{ilkj})
\langle\Phi^{\rm HF}\vert a_{k\sigma'}^\dagger
a^{\phantom{\dagger}}_{l\sigma'} \vert\Phi^{\rm HF}\rangle
a_{i\sigma}^\dagger a^{\phantom{\dagger}}_{j\sigma} ~.
\label{resHamil}
\end{array}
\end{equation}
We have to include the effect of correlations on both, the
$N$-particle ground-state energy, $E_0$, and the (excited)
$N+1$-particle energy $E_{\ul{k}m\sigma}$.

The first problem has already been treated in
Ref.~\cite{BorrmannFulde}. The authors made a variational ansatz for the
ground-state wave function in order to include correlation effects.
We want to re-derive their results by directly making use of the
cumulant representation of the ground-state energy,
Eq.~(\ref{gsenergy}),
\begin{equation}
E_0 = (H\vert \Omega) ~.
\label{gsenergysc}
\end{equation}
It is impossible to determine the wave operator $\Omega$ exactly.
Instead, we make an ansatz for $\Omega$ in the Liouville space taking
into account only the most important processes which are induced by
$H_{\rm res}$. Within the Local Ansatz method they are described by
the following operator~\cite{BorrmannFulde}
\begin{equation}
S^\eta_{IJ} = \ds \frac{1}{4} \sum_{\sigma,\sigma'}
A^\dagger_{J\sigma'} B^{\phantom{\dagger}}_{J\sigma'}
A^\dagger_{I\sigma} B^{\phantom{\dagger}}_{I\sigma}  ~.
\label{defeta}
\end{equation}
$S^\eta_{IJ}$ excites electrons in bonds $I$ and $J$ from a bonding
into an antibonding state, thus changing the charge distribution in
these bonds. This reflects the van-der-Waals interaction between the
bonds $I$ and $J$, as well as the interaction within one bond ($I=J$).
Our ansatz for the wave operator $\Omega$ in Eq.~(\ref{gsenergysc})
reads [see Eqs.~(\ref{linansatz}) and (\ref{eqset})]:
\begin{equation}
\Omega = 1 + \sum_{IJ} \eta^{(0)}_{IJ} S^\eta_{IJ} ~.
\label{omegaans0}
\end{equation}
The coefficients $\eta^{(0)}_{IJ}$ are determined from
\begin{equation}
0 = (S^\eta_{IJ}\vert H \Omega) ~.
\label{omegaeq0}
\end{equation}
Thus we are lead to the same result for the ground-state energy as in
Ref.~\cite{BorrmannFulde}, which we refer to for the further
evaluation of the expectation values.

This procedure to calculate the ground-state energy is advantageous
because it applies for excited states, too, as is shown below.

The energies $E_{\ul{k}m\sigma}$ which we consider next are, in fact,
excited energies of the $N+1$-particle system. According to
Sec.~\ref{sec:theory}, they are given by
\begin{equation}
E_{\ul{k}m\sigma} = ( H\vert \Omega)_{\ul{k}m\sigma} ~,
\label{exenergysc}
\end{equation}
where the expectation values are taken with respect to an excited
$N+1$-particle state $\vert\Phi^{\rm HF}_{\ul{k}m\sigma}\rangle$.
Again, we make an ansatz for the wave operator $\Omega$. In this case,
however, there are two different correlation effects which we have to
take into account. They have already been discussed in detail in
Ref.~\cite{BorrmannFulde}. The first process is the induced
dipole-dipole interaction as it was present in the $N$-particle ground
state above. It is described by the operators $S^\eta_{IJ}$,
Eq.~(\ref{defeta}). The second process is the formation of a
polarization cloud around the added particle. It is described by
\begin{equation}
S^\pi_{IJ} = \ds \frac{1}{4} \sum_{\sigma,\sigma'}
A^\dagger_{J\sigma'} B^{\phantom{\dagger}}_{J\sigma'}
A^\dagger_{I\sigma} A^{\phantom{\dagger}}_{I\sigma} ~.
\end{equation}
This is seen from
\begin{equation}
S^\pi_{IJ} \vert\Phi^{\rm HF}_{\ul{k}m\sigma} \rangle =
\frac{\alpha_{\ul{k}m\sigma}(I)}{4} \sum_{\sigma'}
A^\dagger_{J\sigma'} B^{\phantom{\dagger}}_{J\sigma'}
A^\dagger_{I\sigma}  \vert\Phi^{\rm HF}\rangle
\label{defpi}
\end{equation}
where the extra electron in bond $I$ induces a transition of an
electron in bond $J$ from a bonding to an antibonding state resulting
in a polarization of bond $J$.

For the wave operator $\Omega$ in Eq.~(\ref{exenergysc}) we make the
following ansatz
\begin{equation}
\Omega = 1
+ \sum_{IJ} \eta_{IJ}^{\phantom{\eta}}(\ul{k}m\sigma) S^\eta_{IJ}
+ \sum_{IJ} \pi_{IJ}^{\phantom{\eta}}(\ul{k}m\sigma) S^\pi_{IJ}~.
\label{omegaansk}
\end{equation}
Note that the coefficients describing the dipole-dipole interaction,
$\eta_{IJ}$, do not necessarily equal $\eta_{IJ}^{(0)}$. Furthermore,
the coefficients $\eta_{IJ}(\ul{k}m\sigma)$ and
$\pi_{IJ}(\ul{k}m\sigma)$ may depend on the quantum numbers $\ul{k}$,
$m$, and $\sigma$. They are determined from
\begin{eqnarray}
0 & = & \left( S^\eta_{KL} \vert H \Omega \right)_{\ul{k}m\sigma}
\label{omegaeqetak}\\
0 & = & \left( S^\pi_{KL} \vert H \Omega \right)_{\ul{k}m\sigma} ~.
\label{omegaeqpik}
\end{eqnarray}
This linear system of equations for $\eta_{IJ}(\ul{k}m\sigma)$ and
$\pi_{IJ}(\ul{k}m\sigma)$ leads via Eqs.~(\ref{omegaansk}) and
(\ref{exenergysc}) to improved energies of the $N+1$-particle
system. Subtracting the ground-state energy $E_0$ of the $N$-particle
system, Eq.~(\ref{gsenergysc}) with Eq.~(\ref{omegaans0}), we arrive
at quasiparticle energies $\epsilon_{\ul{k}m\sigma}$, which include
the main effect of electron correlations.

Obviously, the case of one added hole can be treated in close analogy.
This will yield the valence bands.

\section{Discussion}
\label{sec:discussion}
In this section we want to compare the results with previous work done
in Ref.~\cite{BorrmannFulde}. There the same model is treated as here.
Starting from a HF calculation, the authors make a variational ansatz
for both, ground and excited states. The correlation operators are the
same as in Eqs.~(\ref{defeta}) and (\ref{defpi}). The authors consider
the quasiparticle energy directly and, by making use of a
linked-cluster theorem, they introduce cumulants ensuring
size-consistency. Finally, they derive equations for the variational
parameters $\eta_{IJ}$ and $\pi_{IJ}$ by minimizing the energy. The
final equations look similar to the Eqs.~(\ref{omegaeqetak}) and
(\ref{omegaeqpik}) derived here, the main difference being that
$A_{\ul{k}m\sigma}^\dagger$ is subjected to cumulant ordering there,
while it is included in the wave function here and, thus, not
subjected to cumulant ordering. This difference is not surprising:
When the quasiparticle energy is calculated directly, only processes
which are connected to the extra particle contribute, e.g., the
blocking of ground-state correlations in the neighborhood of the extra
particle. This is ensured by subjecting $A^\dagger_{\ul{k}m\sigma}$ to
cumulant ordering. In this work, however, we calculate the energy of
the $N$- and $N\pm 1$-particle system separately. Therefore, we must
also include processes which are not connected to the extra particle,
e.g., ground-state correlations far apart from the extra particle.

By exploiting the properties of cumulants we may rewrite the
expectation values in Eqs.~(\ref{omegaeqetak}) and (\ref{omegaeqpik})
such that they are taken with respect to $\vert\Phi_0\rangle$ only,
thus subjecting $A^\dagger_{\ul{k}m\sigma}$ to cumulant ordering.
Taking the energy difference between $E_{\ul{k}m\sigma}$ and $E_0$, we
arrive at essentially the same quasiparticle energies as in
Ref.~\cite{BorrmannFulde} as will be shown next.

We begin with the expectation values which involve $S^\eta_{IJ}$:
\begin{equation}
\begin{array}{r c l}
(S^\eta_{IJ} \vert H)_{\ul{k}m\sigma} & = &
\langle\Phi^{\rm HF}\vert A^{\phantom{\dagger}}_{\ul{k}m\sigma}
S^{\eta\dagger}_{IJ} H A^\dagger_{\ul{k}m\sigma} \vert \Phi^{\rm HF}
\rangle \\
& = &
\langle\Phi^{\rm HF}\vert A^{\phantom{\dagger}}_{\ul{k}m\sigma}
S^{\eta\dagger}_{IJ} H A^\dagger_{\ul{k}m\sigma} \vert \Phi^{\rm HF}
\rangle^c +  \langle\Phi^{\rm HF}\vert S^{\eta\dagger}_{IJ} H \vert
\Phi^{\rm HF} \rangle \\
& = &
(S^\eta_{IJ} A^\dagger_{\ul{k}m\sigma} \vert
H A^\dagger_{\ul{k}m\sigma})
+ (S^\eta_{IJ} \vert H) ~.
\end{array}
\label{expand1}
\end{equation}
Similarly,
\begin{equation}
(S^\eta_{IJ} \vert H S^\eta_{KL})_{\ul{k}m\sigma} =
\left( S^\eta_{IJ} A^\dagger_{\ul{k}m\sigma} \left\vert
[H-\epsilon^{\rm HF}_{\ul{k}m\sigma}] S^\eta_{KL}
A^\dagger_{\ul{k}m\sigma}\right.\right)
+ (S^\eta_{IJ} \vert H S^\eta_{KL})
\label{expand2}
\end{equation}
since $(A^\dagger_{\ul{k}m\sigma}\vert H A^\dagger_{\ul{k}m\sigma}) =
\epsilon^{\rm HF}_{\ul{k}m\sigma}$. The operators $S^\pi_{IJ}$ are
connected to $A^\dagger_{\ul{k}m\sigma}$. Hence,
\begin{equation}
\begin{array}{r c l}
(S^\pi_{IJ}\vert H)_{\ul{k}m\sigma}
& = & \left(S^\pi_{IJ} A^\dagger_{\ul{k}m\sigma} \left\vert H
A^\dagger_{\ul{k}m\sigma} \right.\right)
\\
(S^\pi_{IJ}\vert H S^\eta_{KL})_{\ul{k}m\sigma}
& = & \left(S^\pi_{IJ} A^\dagger_{\ul{k}m\sigma}\left\vert H
S^\eta_{KL} A^\dagger_{\ul{k}m\sigma}\right.\right) \\
(S^\pi_{IJ}\vert H S^\pi_{KL})_{\ul{k}m\sigma}
& = & \left(S^\pi_{IJ} A^\dagger_{\ul{k}m\sigma}\left\vert
[H-\epsilon^{\rm HF}_{\ul{k}m\sigma}]
S^\pi_{KL} A^\dagger_{\ul{k}m\sigma}\right.\right) ~.
\label{expand3}
\end{array}
\end{equation}

When we insert Eqs.~(\ref{expand1})--(\ref{expand3}) into
Eq.~(\ref{omegaeqetak}) we find
\begin{equation}
\begin{array}{r c l}
0 & = &
\ds \left(S^\eta_{IJ} A^\dagger_{\ul{k}m\sigma} \left\vert H
A^\dagger_{\ul{k}m\sigma}\right.\right) + (S^\eta_{IJ} \vert H) \\
&& \ds + \sum_{KL} \eta_{KL}^{\phantom{\eta}}(\ul{k}m\sigma)
\left\{
\left( S^\eta_{IJ} A^\dagger_{\ul{k}m\sigma} \left\vert
[H-\epsilon^{\rm HF}_{\ul{k}m\sigma}]
S^\eta_{KL} A^\dagger_{\ul{k}m\sigma} \right.\right)
+ (S^\eta_{IJ} \vert H S^\eta_{KL}) \right\} \\
&& \ds + \sum_{KL} \pi_{KL}^{\phantom{\eta}}(\ul{k}m\sigma)
\left(S^\eta_{IJ} A^\dagger_{\ul{k}m\sigma}\left\vert H S^\pi_{KL}
A^\dagger_{\ul{k}m\sigma}\right.\right) ~.
\end{array}
\end{equation}
The individual expectation values differ in their behavior as
$N\to\infty$. While the cumulants involving
$A^\dagger_{\ul{k}m\sigma}$ turn out to be of order $N^{-1}$, the
others are ${\rm O}(N^0)$. To leading order in $N$
\begin{equation}
0 = \ds (S^\eta_{IJ} \vert H)
+ \sum_{KL} \eta_{KL}^{\phantom{\eta}}(\ul{k}m\sigma)
(S^\eta_{IJ} \vert H S^\eta_{KL}) ~.
\end{equation}
Comparing with Eqs.~(\ref{omegaans0}) and (\ref{omegaeq0}) we
find that $\eta_{KL}(\ul{k}m\sigma)$ is the same as for the
$N$-particle system (in leading order in $N$):
\begin{equation}
\eta^{\phantom{0}}_{IJ}(\ul{k}m\sigma) =
\eta^{(0)}_{IJ} + {\rm O}(N^{-1}) ~.
\label{equal}
\end{equation}
This result was obtained in Ref.~\cite{BorrmannFulde}, too.

Next we consider Eq.~(\ref{omegaeqpik}), which determines the
coefficients $\pi_{IJ}$. They describe the formation of the
polarization cloud around the added particle. Inserting again
Eqs.~(\ref{expand1})--(\ref{expand3}) we find
\begin{equation}
\begin{array}{r c l}
0 & = &
\ds \left(S^\pi_{IJ} A^\dagger_{\ul{k}m\sigma} \left\vert H
A^\dagger_{\ul{k}m\sigma}\right.\right)
+ \sum_{KL} \eta_{KL}^{\phantom{\eta}}(\ul{k}m\sigma)
\left(S^\pi_{IJ} A^\dagger_{\ul{k}m\sigma}\left\vert
H S^\eta_{KL} A^\dagger_{\ul{k}m\sigma}\right.\right) \\
&& +\ds \sum_{KL} \pi_{KL}^{\phantom{\eta}}(\ul{k}m\sigma)
\left(S^\pi_{IJ} A^\dagger_{\ul{k}m\sigma}\left\vert
[H - \epsilon^{\rm HF}_{\ul{k}m\sigma}]
S^\pi_{KL} A^\dagger_{\ul{k}m\sigma}\right.\right) ~.
\end{array}
\label{pikms1}
\end{equation}
The only difference to Ref.~\cite{BorrmannFulde} is that instead of
the exact quasiparticle energy $\epsilon_{\ul{k}m\sigma}$ the {\em
Hartree-Fock} energy $\epsilon^{\rm HF}_{\ul{k}m\sigma}$ enters
Eq.~(\ref{pikms1}). This is due to a different introduction of
cumulants and different approximations thereafter: As already
mentioned above, the energies of the $N$- and the $N\pm 1$-particle
system are calculated separately here, whereas the quasiparticle
energy is calculated directly in Ref.~\cite{BorrmannFulde}.

Finally, the quasiparticle energies are given by
\begin{equation}
\begin{array}{r c l}
\lefteqn{\epsilon_{\ul{k}m\sigma} = E_{\ul{k}m\sigma} - E_0
= \left(H\vert\Omega\right)_{\ul{k}m\sigma} -
\left(H\vert\Omega\right)} \\
& = & \ds
\left( H \left\vert 1 +
\sum_{IJ} \eta_{IJ}^{\phantom{\eta}}(\ul{k}m\sigma) S^\eta_{IJ} +
\sum_{IJ} \pi_{IJ}^{\phantom{\eta}}(\ul{k}m\sigma) S^\pi_{IJ}
\right.\right)_{\ul{k}m\sigma}
- ~ \left( H \left\vert 1+\sum_{IJ} \eta^{(0)}_{IJ}
S^\eta_{IJ} \right.\right)~.
\end{array}
\end{equation}
Applying Eq.~(\ref{equal}) this can be rewritten in the form
\begin{equation}
\epsilon_{\ul{k}m\sigma} = \ds
\epsilon^{\rm HF}_{\ul{k}m\sigma} +
\sum_{IJ} \eta_{IJ}^{(0)}
\left(A^\dagger_{\ul{k}m\sigma}\left\vert HS^\eta_{IJ}
A^\dagger_{\ul{k}m\sigma}\right.\right)
+ \sum_{IJ} \pi_{IJ}^{\phantom{\eta}}(\ul{k}m\sigma)
\left( A^\dagger_{\ul{k}m\sigma}\left\vert HS^\pi_{IJ}
A^\dagger_{\ul{k}m\sigma} \right.\right)
\end{equation}
The same expression for the quasiparticle energy was derived in
Ref.~\cite{BorrmannFulde}. By using $\eta_{IJ}^{(0)}$ from a
ground-state calculation and determining $\pi_{IJ}(\ul{k}m\sigma)$
from Eq.~(\ref{pikms1}) the energy bands for diamond are obtained.
The energy gap is found to be only 50\% of the HF energy gap and the
widths of the valence and conduction bands are considerably reduced
from their HF values~\cite{BorrmannFulde}.

\section{Summary and Conclusions}
\label{sec:conclusion}
In this paper, we showed how excitation energies are expressed in
terms of cumulants. This work generalizes the cumulant representation
of the ground-state energy which has been derived
recently~\cite{BeckerFulde}. Introducing cumulants explicitly implies
that all subsequent approximations are {\em a priori} size consistent.
We would like to emphasize that these representations of ground-state
and excitation energies hold irrespectively of the particular
splitting of the Hamiltonian $H$ into a solvable part $H_0$ and the
remaining part $(H-H_0)$, which has to be treated approximately. This
contrasts the standard diagrammatic approaches, in which $H_0$ has to
be a one-particle operator. Therefore, forming the cumulant can be
viewed as considering ``linked diagrams'' only in the case of weakly
correlated systems and as their generalization in the case of strongly
correlated systems for which one-particle wave functions would be a
poor starting point.

To give an application of the theory, we calculated the quasiparticle
energy bands of a semiconductor. We made two separate calculations,
one for the ground-state energy of the charge-neutral system and one
for the (excited) system where one electron (hole) with momentum $k$
is added. With the help of the new representation of excitation
energies it was possible to perform both calculations within the same
formalism, i.e., to apply the same approximations in both cases.
Taking the difference of the two energies, we obtained the
quasiparticle energy. Finally we discussed the connection of the
present calculations with a variational treatment of this
problem~\cite{BorrmannFulde} and essentially reproduced the results
obtained there.

\acknowledgements
We would like to thank Professor Dr.\ K.\ W.\ Becker and Dipl.\ Phys.\
K.\ Fischer for several valuable discussions, from which this work
profited.

\end{document}